\documentclass[useAMS,usenatbib]{mn2e}

\usepackage{times}
\usepackage{graphicx}

\def\pcm3{{\rm\thinspace cm^{-3}}}

\def\n_h{{\rm n_{H}}}

\def\NH1{{$N_{\rm HI}~$}}

\def\ga{{\rm\thinspace gauss}}

\def\Msun{\hbox{$\rm\thinspace M_{\odot}$}}

\def\approxlt{\mathrel{\hbox{\rlap{\lower .5ex \hbox {$\sim$}}
        \raise .15 ex \hbox{$<$}}}}
\def\approxgt{\mathrel{\hbox{\rlap{\lower .5ex \hbox {$\sim$}}
        \raise .15 ex \hbox{$>$}}}}

\def\la{\mathrel{\hbox{\rlap{\hbox{\lower4pt\hbox{$\sim$}}}\hbox{$<$}}}}
\def\ga{\mathrel{\hbox{\rlap{\hbox{\lower4pt\hbox{$\sim$}}}\hbox{$>$}}}}
\newbox\grsign \setbox\grsign=\hbox{$>$} \newdimen\grdimen
\grdimen=\ht\grsign
\newbox\simlessbox \newbox\simgreatbox \newbox\simpropbox
\setbox\simgreatbox=\hbox{\raise.5ex\hbox{$>$}\llap
     {\lower.5ex\hbox{$\sim$}}}\ht1=\grdimen\dp1=0pt
\setbox\simlessbox=\hbox{\raise.5ex\hbox{$<$}\llap
     {\lower.5ex\hbox{$\sim$}}}\ht2=\grdimen\dp2=0pt
\setbox\simpropbox=\hbox{\raise.5ex\hbox{$\propto$}\llap
     {\lower.5ex\hbox{$\sim$}}}\ht2=\grdimen\dp2=0pt
\def\simgreat{\mathrel{\copy\simgreatbox}}

\title[]{New stellar members of the Coma Berenices open star cluster}
\author[S. L. Casewell et al.]{S. L. Casewell\thanks{E-mail:
slc25@star.le.ac.uk}, R. F. Jameson, P. D. Dobbie\\ 
Department of Physics and Astronomy, University of Leicester, University Road, Leicester LE1 7RH, UK\\}

\begin{document}

\date{ }

\pagerange{\pageref{firstpage}--\pageref{lastpage}} \pubyear{2005}

\maketitle

\label{firstpage}

\begin{abstract}
We present the results of a survey of the Coma Berenices open star cluster (Melotte 111), 
undertaken using proper motions from the USNO-B1.0 and photometry from the 2MASS Point 
Source catalogues. We have identified 60 new candidate members with masses in the range 
1.007$\Msun$$<$M$<$0.269$\Msun$. 
For each we have estimated a membership probability by 
extracting control clusters from the proper motion vector diagram. All 60 are found to
have greater than 60 per cent  probability of being clusters more than doubling the
number of known cluster members.
The new luminosity function for the cluster peaks at bright magnitudes, but is rising at K$\approx$ 12, 
indicating that it is likely lower mass members may exist. The mass function also supports this hypothesis.

\end{abstract}


\section{Introduction}

During the last two decades there have been numerous deep surveys of young 
nearby open clusters focusing on the detection of very low mass stellar and 
substellar members (e.g. Jameson \& Skillen 1989, Lodieu et al. 2005). Since 
these objects fade during their evolution, in these 
environments they are comparatively luminous, placing them comfortably within
the reach of 2/4m class telescopes. Furthermore, numerical simulations 
suggest that in clusters with ages less than $\sim$200Myrs, dynamical evolution
should not yet have led to the evaporation of a large proportion of these
members (e.g. de la Fuente Marcos \& de la Fuente Marcos 2000).

For these same reasons, there have been relatively few searchs of this type 
undertaken in older open clusters. Indeed, the few, deep, large area surveys of 
the Hyades, the most extensively studied cluster of its type and the closest to
the Sun (46.3pc; Perryman et al. 1998), have led to the identification of only a 
very small number of low mass stellar and substellar members (e.g. Reid 1993, Gizis 
et al. 1999, Moraux, priv. comm.). This finding is loosely in agreement 
with the predictions of N-body simulations which indicate that less than a fifth
of the original population of substellar members remains tidally bound to
a cluster at the age of the Hyades, 625$\pm$50Myrs (Perryman et al. 1998). 
However, despite this rough consistency, until the very low mass populations
of more open clusters of similar age to the Hyades have been studied it 
seems premature to completely exclude other interpretations for the deficit of low 
mass members here e.g. differences in the initial mass function. Furthermore, 
additional investigations of this nature may also be of use in refining N-body 
simulations.

Therefore we have recently embarked on a survey of the Coma Berenices open star 
cluster (Melotte 111, RA = 12 23 00, Dec = +26 00 00, J2000.0) to extend our 
knowledge of it's luminosity function towards 
the hydrogen burning limit. At first glance this seems a prudent choice of target.
It is the second closest open cluster to the Sun. Hipparcos measurements place
it at d=89.9$\pm$2.1pc (van Leeuwen 1999), in agreement with older ground based 
estimates (e.g. d=85.4$\pm$4.9pc, Nicolet 1981). Furthermore, foreground extinction 
along this line of sight is low, E(B-V)$\approx$0.006$\pm$0.013 (Nicolet 1981). 
The metalicity of the cluster is relatively well constrained. Spectroscopic 
examination of cluster members reveals it to be slightly metal poor with 
respect to the Sun. For example, Cayrel de Strobel (1990) determine [Fe/H]
=-0.065$\pm$0.021 using a sample of eight F, G and K type associates, whereas 
Friel \& Boesgaard  (1992) determine [Fe/H]=-0.052$\pm$0.047 from fourteen F 
and G dwarf members. While estimates of the age of Melotte 111 vary considerably 
from 300Myrs to 1Gyr (e.g. Tsvetkov, 1989),  more recent determinations, based on 
fitting model isochrones to the observed cluster sequence, are bunched around 
400-500Myrs (e.g. Bounatiro \& Arimoto 1993, Odenkirchen 1998). Thus the Coma Berenices 
open cluster is probably marginally younger the Hyades.

However, Melotte 111 is projected over a large area of sky ($\sim$100 sq. deg.)
and contains considerably fewer bright stellar members than the Hyades. For example,
Odenkirchen (1998) determines the cluster tidal radius to be $\sim$5-6pc but finds 
only 34 kinematic members down to V=10.5 within a circular area of radius 5$^{\circ}$ 
centred on the cluster. He estimates the total mass of Melotte 111 to lie in the range 
30-90M$_{\odot}$, which can be compared to estimates of 300-460M$_{\odot}$ for the 
mass of the Hyades (e.g. Oort 1979, Reid 1992). 
Additionally, the small proper motion ($\mu_{\alpha}$=-11.21 $\pm$0.26 mas yr$^{-1}$, 
$\mu_{\delta}$=-9.16 $\pm$0.15 mas 
yr$^{-1}$; van Leeuwen 1999) means that proper motion alone is not a suitable means by 
which to discriminate the members of Melotte 111 from the general field population. 
Fortunately, the convergent point for the cluster is sufficiently distant at $\alpha$ = 6 40 31.2, 
$\delta$ = -41 33 00 (J2000)(Masden et al 2002), that we can expect all the cluster members to have essentially 
the same 
proper motion.

In the first detailed survey of Melotte 111, Trumpler (1938) used proper 
motion, spectrophotometric and radial velocity measurements to identify 37 
probable members, m$_{\rm P}$$<$10.5, in a circular region of 7$^{\circ}$ diameter 
centered on the cluster. A significant additional number of fainter candidate members 
were identified by Artyukhina (1955) from a deep (m$_{\rm P}$$<$15) proper motion 
survey of $\sim$7 sq. degrees of the cluster. Argue \& Kenworthy (1969) performed 
a photographic UBVI survey of a circular field, 3.3$^{\circ}$ in diameter, to a limiting
depth of m$_{\rm P}$=15.5. They rejected all but 2 of her candidates with  
m$_{\rm P}$$>$11 but identified a further 2 faint objects with photometry and proper 
motion which they deemed to be consistent with cluster membership. Subsequently, De 
Luca \& Weis (1981) obtained photoelectric photometry for 88 objects (V$>$11), 
drawn from these two studies. They concluded that only 4 stars, 3 of which 
were listed in Argue \& Kenworthy as probable members, had photometry and astrometry
consistent with association to Melotte 111.   
 
More recently, Bounatiro (1993) has searched an area of 6$^{\circ}$$\times$6$^{\circ}$ 
centered on Melotte 111, using the AGK3 catalogue, and has identified 17 new candidate
members (m$_{\rm P}$$<$12). However, despite Randich et al. (1996) identifying 12 new 
potential low mass members (V$\approx$11.5-16) from a ROSAT PSPC survey of the cluster,
a detailed follow-up study has concluded that none of these are associates of Melotte 
111 (Garcia-Lopez et al. 2000). Odenkirchen et al. (1998) have used the Hipparcos and 
ACT catalogues to perform a comprehensive kinematic and photometric study of 1200 sq. 
degrees around the cluster center complete to a depth of V$\approx$10.5. They find a 
total of $\sim50$ kinematic associates which appear to be distributed in a core-halo 
system. The core, which is elliptical in shape with a semi-major axis of 1.6$^{\circ}$,
is dominated by the more massive members while the halo contains proportionately more 
low mass associates. Odenkirchen et al. also find evidence of an extra-tidal
'moving group' located in front of the cluster in the context of its motion around 
the Galaxy.  However, from a subsequent spectroscopic study, Ford et al. (2001),
concluded that approximately half of the moving group were not associated
with the cluster.

The ready availability of high quality astrometric and photometric survey catalogues 
(e.g. USNO-B1.0,  2MASS) have made a new deeper, wide area survey of the Coma Berenices 
open star cluster a tractable undertaking. In this paper we report on our efforts to 
use the USNO-B1.0 and 2MASS Point Source Catalogues to search for further candidate low 
mass members of Melotte 111. We identify 60 new candidates with proper motions and photometry 
consistent with cluster membership. State-of-the-art evolutionary models indicate some of 
these objects have masses of only M$\approx$0.269$\Msun$.

\section{The present survey}

In ideal cases, membership of nearby open clusters can be ascertained by either photometry or
proper motion. In the former approach, associates of a cluster are generally identified 
by insisting that they have magnitudes and colours which, within uncertainties, sit them
on or close to an appropriate model isochrone (e.g. Bouvier et al. 1998). In the latter 
approach, as measurement errors usually dominate over the cluster velocity dispersion, 
selection of members simply requires that objects have astrometric motions consistent with
known associates of a cluster. When combined, these two methods can yield candidates with 
a very high probability of cluster membership (e.g. Moraux et al. 2001)

However, as discussed in \S 1, the proper motion of Melotte 111 is comparatively small 
and not suitable alone for the discrimation of cluster members. For example, members 
of the Pleiades open cluster, which can be readily identified astrometrically, have proper
motions of $\mu_{\alpha}$=19.14$\pm$0.25 mas yr$^{-1}$, $\mu_{\delta}$=-45.25 $\pm$0.19 
mas yr$^{-1}$ (van Leeuwen 1999). Nevertheless, because of the large epoch difference of 
approximately 50 years between the two Palomar Sky Surveys, if we restrict the search to 
R$<$18.0, the proper motion measurements available in the USNO-B1.0 catalogue are sufficiently 
accurate that they can be used in conjunction with suitable photometry to identify candidates 
with a substantial probability of being members of Melotte 111. In principle the USNO-B1.0 
catalogue provides B, R and I photometry but the significant uncertainties in the photographic
magnitudes ($\sim$0.3 mags) severely limit its usefulness for this work (Monet et al. 2003). Therefore, to 
compliment the astrometry, we choose to use  J, H and K$_{\rm S}$ photometry from the 2MASS
Point Source Catalogue which has S/N$\simgreat$10 down to  J=15.8, H=15.1 and K$_{\rm S}$=14.3 
(Skrutskie et al. 1997).

\begin{figure}
\scalebox{0.325}{{\includegraphics{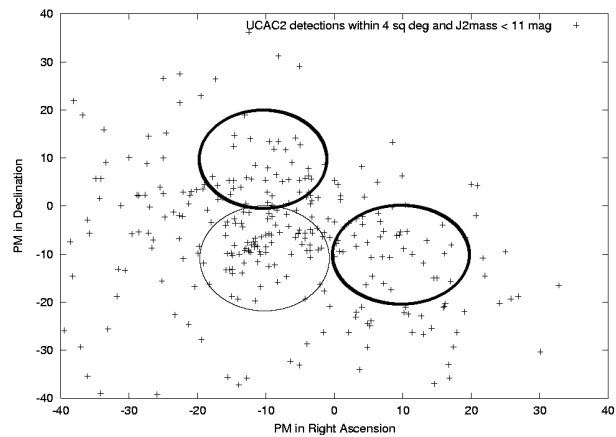}}}
\caption{Plot of the UCAC (Zacharias et al. 2004) proper motions for 4 square degrees of the cluster centre. The thin line borders the cluster proper motion selection, and the thick line, the contols. The shallower UCAC catalogue is used here for illustration as the USNO B1.0 catalogue only provides quantized values of 2 mas yr$^{-1}$. 
}
\end{figure}

\vspace{0.1cm} 

The survey was conducted as follows:-
\begin{enumerate}
\item A circular area of radius 4 degrees centred on the cluster was extracted from the USNO B1.0 catalogue.\\
\item Stars were selected according to the criterion ($\mu_{\alpha}$ - X)$^{2}$ + ($\mu_{\delta}$ - Y)$^{2}$ $<$ 100,  where X = -11.21 and Y = -9.16, i.e. to lie within 10 mas yr$^{-1}$ of the Hipparcos determined value for the proper motion of Melotte 111.  This procedure was initially repeated for 
X  = -11.21,  Y  = +9.16 and X = +11.21,  Y = -9.16, to obtain two control samples (see Figure 1).
Two more control samples were later extracted from two further circular regions of USNO B1.0 data. These
 data had a similar 
 galactic latitude to the cluster but were offset by 10$^{\circ}$ from the centre of Melotte 111.  Stars were selected from these latter regions 
by applying the first proper motion criterion above. 
 The known members have errors, or velocity dispersions ammounting to about $\pm$ 2.0 km s$^{-1}$,
(Odenkirchen et al 1998).
This corresponds to a proper motion of $\pm$ 4.8 mas yr $^{-1}$, and the USNO B1.0 catalogue astrometry errors, (see Tables 1 and 2) are small - typically less than $\pm$ 5 mas yr $^{-1}$
for our stars.  Note the USNO B1.0 proper motion errors are quantized in units of 1 mas yr$^{-1}$ and a zero error thus indicates an error of less than 0.5 mas yr $^{-1}$.
Thus if we selected a bounding circle of 10 mas yr, and the total quadratically added error is 7 mas yr$^{-1}$, we have selected all stars to a completeness level of 1.4 $\sigma$, which means our survey is complete to $\approx$ 90\%.\\
\item Stars passing the above sets of criteria were cross-referenced against the 2MASS point source catalogue using a match radius 
 of 2 arcseconds.\\
\item Subsequently the sample of candidate cluster members and the four control samples were plotted in  K$_{\rm S}$, J-K$_{\rm S}$ colour-magnitude diagrams as illustrated in 
Figures 2a and 2b respectively.

\end{enumerate}

\begin{figure}

\scalebox{0.45}{{\includegraphics{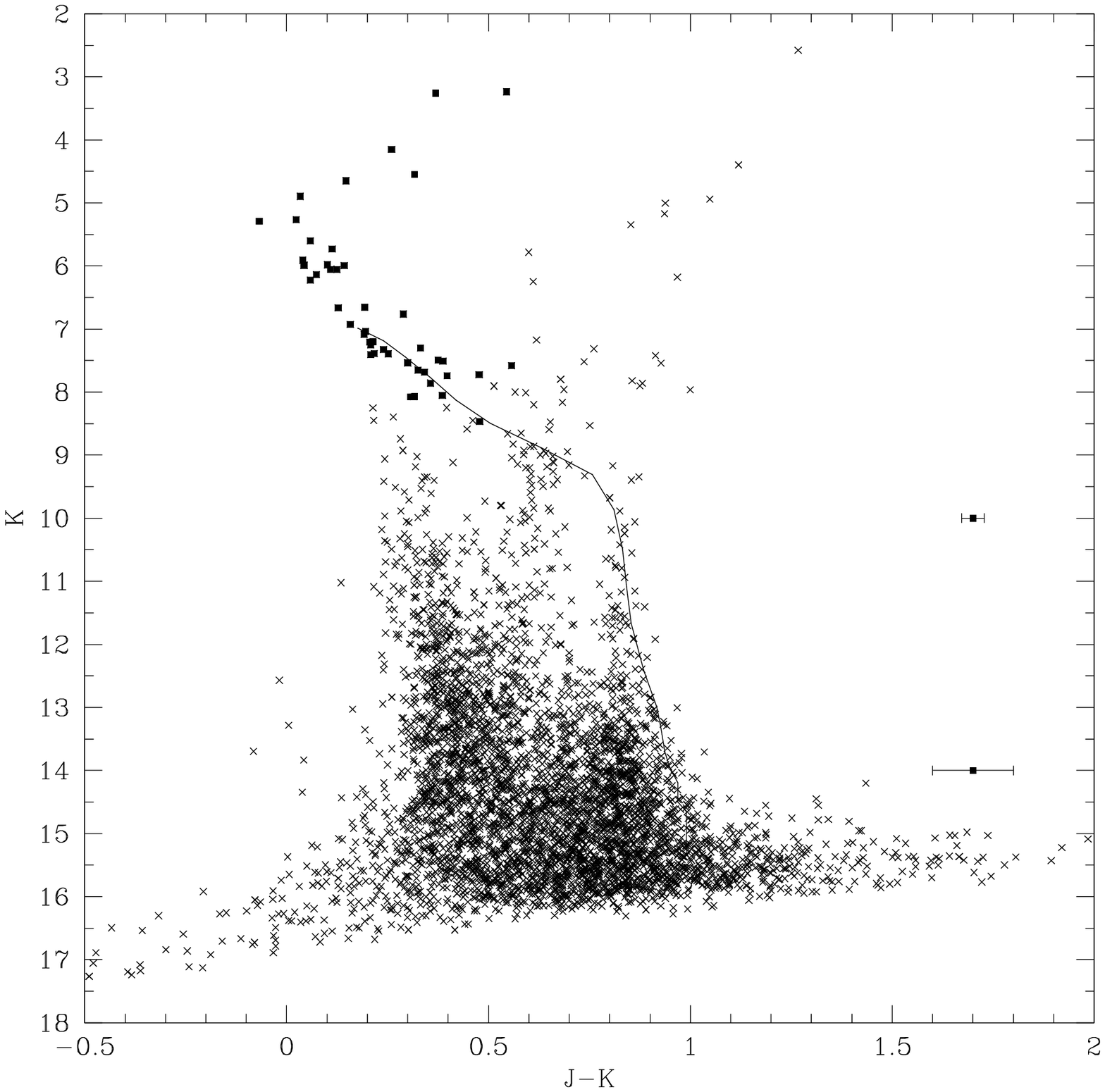}}}
\scalebox{0.45}{{\includegraphics{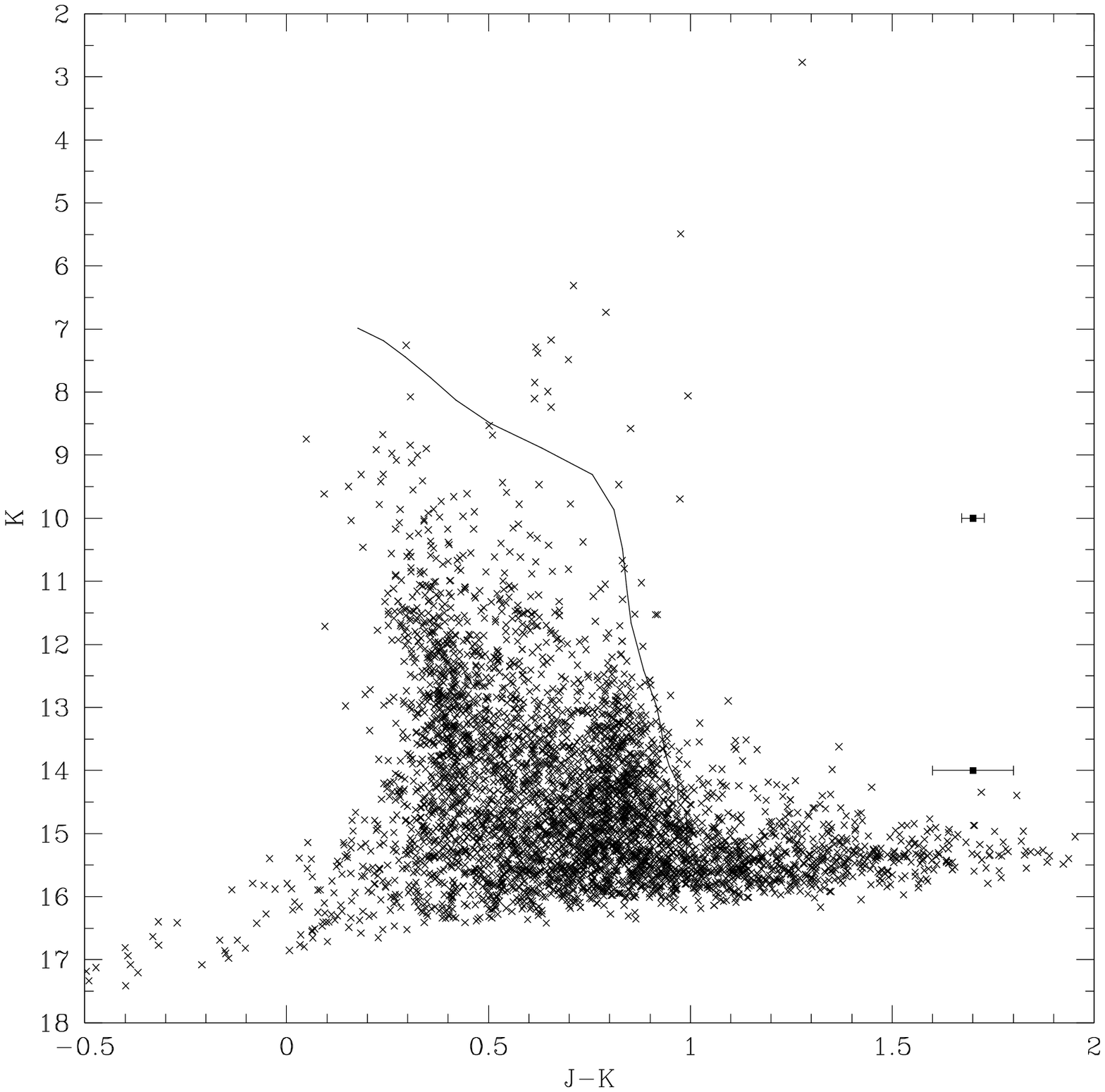}}}
\caption{The CMD for the cluster (top) and a control sample with $\mu_{\alpha}$=+11.21  $\mu_{\delta}$=-9.16 (bottom). Previously known members of the cluster 
are highlighted in the upper plot (solid squares). A 0.4Gyr NEXTGEN isochrone is overplotted from Baraffe et al (1998) (solid line). This was converted into the 2MASS 
system using the transforms of Carpenter(2001).}
\end{figure}

\begin{table*}
\begin{center}
\caption{Name,coordinates,proper motion measurements, R,I,J,H and K magnitudes for the known cluster members as detailed in Bounatiro and Arimoto (1992) and Odenkirchen et al. (1998) for the area we surveyed. The masses were calculated by linearly interpolating the models of Girardi et al(2002) M$\geq$1 $\Msun$
or Baraffe et al(1998),M$<$1 $\Msun$.}
\begin{tabular}{l c c r r r r c c c c c c c}
\hline
Name &RA & Dec.& $\mu_{\alpha}$& $\mu_{\delta}$  & error$\mu_{\alpha}$ &error  $\mu_{\delta}$ & R & I  &J & H & K$_{\rm S}$ &Mass \\
&\multicolumn{2}{|c|}{J2000.0}& \multicolumn{4}{|c|}{mas yr$^{-1}$} &&&&&&$\Msun$ \\
\hline
BD+26 2337 &12 22 30.31 & +25 50 46.1 & -12.0 & -10.0 & 0.0 & 0.0  & 4.54 & 4.28 & 3.78 & 3.40 & 3.23&2.709\\
BD+28 2156 &12 51 41.92 & +27 32 26.5 & -10.0 & -10.0 & 0.0 & 0.0  & 4.55 & 4.20 & 3.62 & 3.36 & 3.26&2.709\\
BD+28 2115 &12 26 24.06 & +27 16 05.6 & -14.0 & -10.0 & 0.0 & 0.0  & 4.79 & 4.65 & 4.40 & 4.23 & 4.14&2.685\\
BD+24 2464 &12 29 27.04 & +24 06 32.1 & -18.0 &   0.0 & 0.0 & 0.0  & 5.25 & 5.02 & 4.86 & 4.57 & 4.54&2.595\\
BD+27 2134 &12 26 59.29 & +26 49 32.5 & -14.0 & -10.0 & 0.0 & 0.0  & 4.91 & 4.88 & 4.79 & 4.72 & 4.64&2.568\\
BD+26 2344 &12 24 18.53 & +26 05 54.9 & -14.0 & -14.0 & 0.0 & 0.0  & 5.11 & 5.08 & 4.93 & 4.94 & 4.89&2.496\\
BD+25 2517 &12 31 00.56 & +24 34 01.8 & -12.0 & -10.0 & 0.0 & 0.0  & 5.41 & 5.39 & 5.29 & 5.30 & 5.26&2.363\\
BD+26 2354 &12 28 54.70 & +25 54 46.2 & -24.0 & -16.0 & 0.0 & 0.0  & 5.24 & 5.28 & 5.22 & 5.29 & 5.28&2.356\\
HD 105805  &12 10 46.09 & +27 16 53.4 & -12.0 & -12.0 & 0.0 & 0.0  & 5.92 & 5.86 & 5.65 & 5.66 & 5.60&2.216\\
BD+26 2345 &12 24 26.79 & +25 34 56.8 & -14.0 & -14.0 & 0.0 & 0.0  & 6.59 & 6.46 & 5.84 & 5.77 & 5.73&2.150\\
BD+23 2448 &12 19 19.19 & +23 02 04.8 & -14.0 & -10.0 & 0.0 & 0.0  & 6.14 & 6.04 & 5.94 & 5.96 & 5.90&2.057\\
BD+26 2326 &12 19 02.02 & +26 00 30.0 & -14.0 & -10.0 & 0.0 & 0.0  & 6.36 & 6.27 & 6.08 & 6.00 & 5.98&2.018\\
BD+25 2523 &12 33 34.21 & +24 16 58.7 & -12.0 & -10.0 & 0.0 & 0.0  & 6.19 & 6.13 & 6.03 & 5.98 & 5.98&2.014\\
BD+27 2138 &12 28 38.15 & +26 13 37.0 & -16.0 &  -8.0 & 0.0 & 0.0  & 6.40 & 6.30 & 6.13 & 6.02 & 5.99&2.011\\
BD+26 2343 &12 24 03.46 & +25 51 04.4 & -14.0 & -10.0 & 0.0 & 0.0  & 6.59 & 6.47 & 6.17 & 6.07 & 6.05&1.978\\
BD+26 2353 &12 28 44.56 & +25 53 57.5 & -22.0 & -18.0 & 0.0 & 0.0  & 6.52 & 6.40 & 6.16 & 6.10 & 6.05&1.977\\
BD+29 2280 &12 19 50.62 & +28 27 51.6 & -12.0 & -10.0 & 0.0 & 0.0  & 6.52 & 6.42 & 6.20 & 6.19 & 6.13&1.931\\
BD+26 2352 &12 27 38.36 & +25 54 43.5 & -14.0 & -12.0 & 0.0 & 0.0  & 6.57 & 6.48 & 6.28 & 6.22 & 6.22&1.879\\
BD+30 2287 &12 31 50.55 & +29 18 50.9 & -12.0 & -10.0 & 0.0 & 0.0  & 7.37 & 7.20 & 6.84 & 6.74 & 6.65&1.607\\
BD+25 2495 &12 21 26.74 & +24 59 49.2 & -12.0 & -10.0 & 0.0 & 0.0  & 7.23 & 7.08 & 6.79 & 6.74 & 6.66&1.600\\
BD+26 2347 &12 25 02.25 & +25 33 38.3 & -14.0 &  -8.0 & 0.0 & 0.0  & 7.83 & 7.55 & 7.05 & 6.85 & 6.76&1.540\\
BD+26 2323 &12 17 50.90 & +25 34 16.8 & -12.0 & -12.0 & 0.0 & 0.0  & 7.66 & 7.47 & 7.08 & 6.98 & 6.92&1.451\\
BD+26 2321 &12 16 08.37 & +25 45 37.3 & -12.0 & -10.0 & 0.0 & 0.0  & 7.87 & 7.66 & 7.23 & 7.11 & 7.03&1.399\\
BD+28 2087 &12 12 24.89 & +27 22 48.3 & -12.0 & -12.0 & 0.0 & 0.0  & 7.85 & 7.64 & 7.27 & 7.13 & 7.08&1.380\\
BD+28 2095 &12 16 02.50 & +28 02 55.2 & -24.0 &  -6.0 & 0.0 & 0.0  & 8.03 & 7.80 & 7.41 & 7.22 & 7.20&1.331\\
BD+27 2129 &12 25 51.95 & +26 46 36.0 & -14.0 & -10.0 & 0.0 & 0.0  & 8.09 & 7.86 & 7.41 & 7.30 & 7.20&1.331\\
BD+27 2122 &12 23 41.00 & +26 58 47.7 & -14.0 & -10.0 & 0.0 & 0.0  & 8.10 & 7.87 & 7.46 & 7.33 & 7.25&1.313\\
BD+23 2447 &12 18 36.17 & +23 07 12.2 & -14.0 & -10.0 & 0.0 & 0.0  & 8.39 & 8.13 & 7.63 & 7.38 & 7.30&1.294\\
BD+28 2109 &12 21 56.16 & +27 18 34.2 & -10.0 & -12.0 & 0.0 & 0.0  & 8.22 & 7.96 & 7.56 & 7.39 & 7.32&1.285\\
HD 107685  &12 22 24.75 & +22 27 50.9 & -12.0 & -10.0 & 0.0 & 0.0  & 8.23 & 8.00 & 7.60 & 7.39 & 7.38&1.263\\
BD+24 2457 &12 25 22.49 & +23 13 44.7 & -14.0 & -10.0 & 0.0 & 0.0  & 8.30 & 8.06 & 7.64 & 7.48 & 7.39&1.261\\
BD+28 2125 &12 31 03.09 & +27 43 49.2 & -16.0 &  -8.0 & 0.0 & 0.0  & 8.27 & 8.01 & 7.61 & 7.46 & 7.40&1.256\\
BD+25 2488 &12 19 28.35 & +24 17 03.2 & -12.0 & -12.0 & 0.0 & 0.0  & 8.69 & 8.40 & 7.86 & 7.55 & 7.49&1.224\\
HD 109483  &12 34 54.29 & +27 27 20.2 & -12.0 & -10.0 & 0.0 & 0.0  & 8.67 & 8.40 & 7.89 & 7.58 & 7.51&1.217\\
BD+25 2486 &12 19 01.47 & +24 50 46.1 & -12.0 & -10.0 & 0.0 & 0.0  & 8.53 & 8.27 & 7.83 & 7.55 & 7.53&1.207\\
BD+27 2130 &12 26 05.48 & +26 44 38.3 &  -8.0 & -14.0 & 0.0 & 0.0  & 9.44 & 9.04 & 8.13 & 7.68 & 7.58&1.193 \\
HD 107399  &12 20 45.56 & +25 45 57.1 & -12.0 &  -8.0 & 0.0 & 0.0  & 8.68 & 8.42 & 7.97 & 7.74 & 7.65&1.171\\
BD+26 2340 &12 23 08.39 & +25 51 04.9 & -12.0 & -10.0 & 0.0 & 0.0  & 8.80 & 8.52 & 8.02 & 7.76 & 7.68&1.160\\
BD+25 2511 &12 29 40.92 & +24 31 14.6 & -10.0 & -10.0 & 0.0 & 0.0  & 9.26 & 8.80 & 8.20 & 7.84 & 7.72&1.148\\
BD+27 2121 &12 23 41.82 & +26 36 05.3 & -16.0 & -12.0 & 0.0 & 0.0  & 8.85 & 8.49 & 8.13 & 7.79 & 7.73&1.143\\
BD+27 2117 &12 21 49.02 & +26 32 56.7 & -12.0 &  -8.0 & 0.0 & 0.0  & 8.89 & 8.57 & 8.21 & 7.86 & 7.85&1.109\\
TYC1991-1087-1 &12 27  48.29& +28 11  39.8& -12.0& -10.0&  0.0&  0.0 &     9.26&  8.94&     8.43&    8.05&   8.05&1.051\\
HD 105863  &12 11 07.38  & +25 59 24.6 &-12.0 &-10.0& 0.0 &0.0 & 9.14  &8.84 &8.38  & 8.11 & 8.07 &1.043\\
BD+30 2281 &12 29 30.02 & +29 30 45.8 & 10.0  &   0.0  &   0.0 &    0.0&  9.41&  9.12& 8.38  &   8.16&    8.07&1.043\\  
BD+36 2312 &12 28 21.11  & +28 02 25.9& -14.0 & -12.0&   0.0&   0.0&     9.92&  9.54 &  8.94&  8.47&     8.46&0.915\\
\hline
\end{tabular}
\end{center}
\end{table*}

A cursory glance at the CMDs in Figure 2 reveals that our method appears to be 
rather successful in finding new associates of Melotte 111. An obvious cluster 
sequence can be seen extending to K$\approx$12 beyond which it is overwhelmed 
by field star contamination. To perform a quantitative selection of candidate 
members we restrict ourselves to K$_{\rm S}$$<$12. We use as a guide to the 
location of the cluster sequence the previously known members and a 400Myr NEXTGEN 
isochrone for solar metalicity (Baraffe et al. 1998), scaled to the Hipparcos 
distance determination for Melotte 111. In the magnitude range where they overlap, 
the previously known cluster members of the single star sequence, congregate 
around the theoretical isochrone (all previously known cluster members are listed in Table 1).
Furthermore, the model isochrone appears to 
continue to follow closely the excess of objects in Figure 2a, relative to Figure 2b, 
suggesting that it is relatively robust in this effective temperature regime.

The location of the theoretical isochrone at J-K$_{\rm S}$$<0.8$ is insensitive to the uncertainties 
in the age of the cluster.  The cluster would have to be much younger to significantly shift the 
isochrone.  The bulk of 
the observed dispersion in the single star sequence here likely stems from the 
finite depth of the cluster ($\sim$0.15 mags). Nevertheless, in this colour range 
we choose to select objects which lie no more than 0.3 magnitudes below the 
theoretical isochrone as this ensures that all previously known cluster members are
recovered (filled squares in Figure 2a). As binary members can lie up to 0.75 magnitudes
above the single star sequence, only objects which also lie no 
more than 1.05 magnitudes above the theoretical sequence are deemed candidate members.  

Redward of J-K$_{\rm S}$=0.8, the main sequence becomes very steep in this CMD. We have 
tried using various combinations of 2MASS and USNO-B1.0 photometry (e.g. colours such as 
R-K) to circumvent this. However, as mentioned previously, the poorer quality of the 
photographic magnitudes provided by the USNO-B1.0 catalogue results in a large amount 
of scatter in optical+IR CMDs rendering them of little use for this work. 
Nonetheless, the finite depth of the cluster and a small error in cluster 
distance determination have a negligible effect in this part of the K$_{\rm S}$, J-K$_{\rm S}$ 
CMD. Based on previous experience gained from our investigations of the low 
mass members of the Pleiades and Praesepe open clusters, we estimate an uncertainty in
the model J-K colour of $\pm$ 0.05 magnitudes. Hence for J-K$_{\rm S}$$>$0.8, we have 
selected all objects which overlap a region 0.1 magnitudes wide in J-K, centered on the 
theoretical isochrone. 

To assess levels of contamination in our list of candidate members, we have  
imposed these same colour selection criteria on the control samples. The 
resulting sequences were divided into one magnitude bins (in K$_{\rm S}$)for J-K$_{\rm S}$$>$0.8, 
and in bins of 0.2 in J-K$_{\rm S}$ for J-K$_{\rm S}$$<$0.8 and the number 
of objects in each for both the cluster and control samples counted. Subsequently, the
membership probablility for each candidate cluster member, $P_{\rm membership}$,
was estimated using equation 1, 

\begin{equation}
P_{\rm membership}=\frac{N_{\rm cluster}-N_{\rm control}}{N_{\rm cluster}}
\label{eqno1}
\end{equation}
where  $N_{\rm cluster}$ is the number of stars in a magnitude bin from the cluster sample
and $N_{\rm control}$ is the mean number of stars in the same magnitude range but drawn from
the control samples. Our list of candidate associates of Melotte 111 is presented in Table 2, 
along with these estimates of membership probability.

We note that there is a slight increase in the level of contamination in the range 
0.55$<$J-K$_{\rm S}$$<$0.7. A similar increase in the number of field stars was seen by
Adams et al.(2002) when studying the Praesepe open cluster, which, like Melotte 111, has 
a relatively high galactic latitude, b=38$^{\circ}$. We believe this is due to K giants
located in the thick disc.

\begin{table*}
\begin{center}
\caption{Coordinates,proper motion measurements, R,I,J,H and K magnitudes. Mass calculated from linearly interpolating the NEXTGEN model and the absolute K magnitude. 
 The probability of membership for each of our 60 possible members.}
\begin{tabular}{c c r r r r r r r r r c c}
\hline
RA & Dec.& $\mu_{\alpha}$& $\mu_{\delta}$  & error$\mu_{\alpha}$ &error  $\mu_{\delta}$  & R & I & J & H & K$_{\rm S}$ &
Mass & Membership\\
\multicolumn{2}{|c|}{J2000.0}& \multicolumn{4}{|c|}{mas yr$^{-1}$} &&&&&&$\Msun$ & probablility \\
\hline
12 24 17.15 & +24 19 28.4 & -18.0 & -14.0 & 0.0 & 0.0 &  9.44 &  9.08 &   8.42 &  7.95 &   7.90 &0.880&0.64\\
12 38 14.94 & +26 21 28.1 &  -6.0 &  -2.0 & 0.0 & 0.0 &  9.57 &  9.13 &   8.56 &  8.07 &   8.00 &0.836&0.64\\
12 23 28.69 & +22 50 55.8 & -14.0 & -10.0 & 0.0 & 0.0 &  9.77 &  9.23 &   8.60 &  8.09 &   8.01 &0.815&0.64\\
12 31 04.78 & +24 15 45.4 &  -8.0 &  -4.0 & 0.0 & 0.0 &  9.76 &  9.19 &   8.81 &  8.28 &   8.20 &0.798&0.79\\
12 27 06.26 & +26 50 44.5 & -12.0 &  -8.0 & 0.0 & 0.0 &  9.59 &  9.31 &   8.64 &  8.33 &   8.25 &1.007&0.94\\
12 27 20.69 & +23 19 47.5 & -14.0 & -10.0 & 0.0 & 0.0 &  9.83 &  9.46 &   8.91 &  8.54 &   8.45 &0.936&0.64\\
12 23 11.99 & +29 14 59.9 &  -2.0 &  -6.0 & 0.0 & 0.0 & 10.29 &  9.84 &   9.13 &  8.58 &   8.47 &0.764&0.79\\
12 33 30.19 & +26 10 00.1 & -16.0 & -10.0 & 0.0 & 0.0 & 10.59 & 10.14 &   9.24 &  8.72 &   8.59 &0.766&0.79\\
12 39 52.43 & +25 46 33.0 & -20.0 &  -8.0 & 0.0 & 0.0 & 10.79 & 10.49 &   9.23 &  8.74 &   8.65 &0.824&0.64\\
12 28 56.43 & +26 32 57.4 & -14.0 &  -8.0 & 0.0 & 0.0 & 10.53 & 10.25 &   9.21 &  8.77 &   8.66 &0.852&0.64\\
12 24 53.60 & +23 43 04.9 &  -4.0 &  -8.0 & 0.0 & 0.0 & 10.73 & 10.34 &   9.39 &  8.86 &   8.82 &0.840&0.64\\
12 28 34.29 & +23 32 30.6 &  -8.0 & -14.0 & 0.0 & 0.0 & 10.40 &  9.97 &   9.47 &  8.93 &   8.86 &0.798&0.79\\
12 33 00.62 & +27 42 44.8 & -14.0 & -14.0 & 0.0 & 0.0 & 10.42 &  9.80 &   9.47 &  8.94 &   8.87 &0.805&0.79\\
12 35 17.03 & +26 03 21.8 &  -2.0 & -10.0 & 0.0 & 0.0 & 10.69 & 10.35 &   9.51 &  9.01 &   8.93 &0.818&0.64\\
12 25 10.14 & +27 39 44.8 &  -6.0 & -12.0 & 0.0 & 0.0 & 10.69 & 10.10 &   9.57 &  9.07 &   8.93 &0.777&0.79\\
12 18 57.27 & +25 53 11.1 & -12.0 & -12.0 & 3.0 & 1.0 & 10.80 & 10.30 &   9.64 &  9.08 &   8.94 &0.728&0.79\\
12 21 15.63 & +26 09 14.1 & -10.0 &  -8.0 & 0.0 & 0.0 & 10.87 & 10.41 &   9.62 &  9.09 &   8.97 &0.773&0.79\\
12 32 08.09 & +28 54 06.5 & -10.0 &  -4.0 & 0.0 & 0.0 & 10.74 & 10.30 &   9.62 &  9.09 &   8.99 &0.785&0.79\\
12 12 53.23 & +26 15 01.3 & -12.0 & -12.0 & 0.0 & 0.0 & 10.54 &  9.72 &   9.58 &  9.11 &   8.99 &0.819&0.64\\
12 23 47.23 & +23 14 44.3 & -12.0 & -16.0 & 0.0 & 0.0 & 10.36 &  9.54 &   9.68 &  9.13 &   9.02 &0.759&0.79\\
12 18 17.77 & +23 38 32.8 &  -6.0 & -14.0 & 0.0 & 0.0 & 10.77 & 10.28 &   9.76 &  9.20 &   9.10 &0.759&0.79\\
12 22 52.37 & +26 38 24.2 &  -8.0 & -10.0 & 0.0 & 0.0 & 11.42 & 11.12 &   9.78 &  9.26 &   9.11 &0.756&0.79\\
12 26 51.03 & +26 16 01.9 & -14.0 &  -2.0 & 0.0 & 0.0 & 11.07 & 10.56 &   9.85 &  9.27 &   9.16 &0.725&0.79\\
12 09 12.44 & +26 39 38.9 & -16.0 &  -6.0 & 0.0 & 0.0 & 10.60 &  9.78 &   9.83 &  9.27 &   9.18 &0.770&0.79\\
12 27 00.81 & +29 36 37.9 &  -4.0 &  -6.0 & 0.0 & 0.0 & 11.07 & 10.70 &   9.80 &  9.33 &   9.20 &0.806&0.79\\
12 23 28.21 & +25 53 39.9 & -10.0 & -12.0 & 1.0 & 1.0 & 11.43 & 11.00 &   9.92 &  9.35 &   9.26 &0.758&0.79\\
12 24 10.37 & +29 29 19.6 &  -6.0 &  -2.0 & 0.0 & 0.0 & 10.77 & 10.27 &  10.06 &  9.50 &   9.33 &0.683&0.79\\
12 34 46.93 & +24 09 37.7 & -12.0 &  -4.0 & 2.0 & 5.0 & 11.38 & 10.87 &  10.06 &  9.53 &   9.39 &0.749&0.79\\
12 15 34.01 & +26 15 42.9 & -12.0 &  -6.0 & 0.0 & 0.0 & 11.08 & 10.26 &  10.13 &  9.57 &   9.47 &0.757&0.79\\
12 26 00.26 & +24 09 20.9 & -10.0&   -4.0 & 2.0 & 2.0 & 13.85 & 12.51 &  10.98 & 10.36 &  10.14 &0.558&0.80\\
12 28 57.67 & +27 46 48.4 & -14.0 &  -2.0 & 3.0 & 2.0 & 13.04 & 11.37 &  10.99 & 10.35 &  10.19 &0.552&0.80\\
12 16 00.86 & +28 05 48.1 & -12.0 &  -4.0 & 2.0 & 2.0 & 14.15 & 11.43 &  11.07 & 10.52 &  10.24 &0.543&0.80\\ 
12 30 57.39 & +22 46 15.2 & -12.0 &  -4.0 & 3.0 & 1.0 & 14.37 & 12.13 &  11.24 & 10.65 &  10.42 &0.514&0.80\\
12 31 57.42 & +25 08 42.5 & -10.0 & -12.0 & 0.0 & 1.0 & 14.22 & 12.94 &  11.40 & 10.79 &  10.55 &0.492&0.80\\
12 31 27.72 & +25 23 39.9 &  -8.0 &  -6.0 & 1.0 & 2.0 & 14.07 & 12.96 &  11.44 & 10.84 &  10.63 &0.478&0.80\\
12 25 55.76 & +29 07 38.3 &  -8.0 & -18.0 & 2.0 & 3.0 & 13.81 & 11.78 &  11.56 & 10.95 &  10.75 &0.460&0.80\\
12 23 55.53 & +23 24 52.3 & -10.0 &  -4.0 & 1.0 & 0.0 & 14.37 & 12.64 &  11.59 & 10.99 &  10.77 &0.455&0.80\\
12 25 02.64 & +26 42 38.4 &  -8.0 &  -4.0 & 0.0 & 3.0 & 14.22 & 11.85 &  11.62 & 11.03 &  10.79 &0.452&0.80\\
12 30 04.87 & +24 02 33.9 & -10.0 &  -8.0 & 3.0 & 0.0 & 14.76 & 13.27 &  11.77 & 11.18 &  10.94 &0.428&0.80\\
12 18 12.77 & +26 49 15.6 &  -8.0&    0.0 & 1.0 & 1.0 & 15.61 & 12.78 &  12.02 & 11.46 &  11.15 &0.392&0.60\\
12 15 16.93 & +28 44 50.0 & -12.0 & -14.0 & 4.0 & 1.0 & 14.11 & 0.0   &  12.00 & 11.35 &  11.17 &0.389&0.60\\
12 23 12.03 & +23 56 15.1 &  -8.0 &  -6.0 & 2.0 & 1.0 & 15.40 & 13.31 &  12.20 & 11.61 &  11.38 &0.352&0.60\\
12 31 00.28 & +26 56 25.1 & -16.0 & -14.0 & 3.0 & 1.0 & 14.04 & 12.52 &  12.25 & 11.56 &  11.39 &0.351&0.60\\
12 33 31.35 & +24 12 09.1 &  -4.0 & -16.0 & 1.0 & 2.0 & 14.76 & 13.45 &  12.29 & 11.59 &  11.40 &0.348&0.60\\
12 16 37.30 & +26 53 58.2 & -10.0 &  -6.0 & 2.0 & 3.0 & 15.13 & 12.94 &  12.23 & 11.68 &  11.42 &0.346&0.60\\
12 24 10.89 & +23 59 36.4 &  -6.0 &  -4.0 & 4.0 & 1.0 & 15.60 & 13.67 &  12.27 & 11.66 &  11.45 &0.340&0.60\\
12 28 38.70 & +25 59 13.0 &  -6.0 &  -2.0 & 1.0 & 2.0 & 14.31 & 13.94 &  12.36 & 11.69 &  11.53 &0.326&0.60\\
12 16 22.84 & +24 19 01.1 & -12.0 &  -2.0 & 2.0 & 4.0 & 14.78 & 13.96 &  12.39 & 11.73 &  11.55 &0.323&0.60\\
12 27 08.56 & +27 01 22.9 & -16.0 &  -2.0 & 2.0 & 3.0 & 14.14 & 12.52 &  12.38 & 11.73 &  11.57 &0.319&0.60\\
12 28 04.54 & +24 21 07.6 & -12.0 & -10.0 & 1.0 & 3.0 & 15.75 & 14.29 &  12.39 & 11.84 &  11.58 &0.318&0.60\\
12 38 04.72 & +25 51 18.5 & -16.0 &  -6.0 & 4.0 & 0.0 & 14.96 & 13.43 &  12.46 & 11.84 &  11.64 &0.308&0.60\\
12 14 19.78 & +25 10 46.6 &  -6.0 & -12.0 & 3.0 & 3.0 & 14.81 & 13.92 &  12.50 & 11.80 &  11.65 &0.305&0.60\\
12 28 50.08 & +27 17 41.7 & -20.0 & -12.0 & 1.0 & 2.0 & 14.46 & 12.84 &  12.55 & 11.84 &  11.70 &0.297&0.60\\
12 36 34.30 & +25 00 38.3 & -14.0 &  -2.0 & 4.0 & 6.0 & 15.59 & 13.43 &  12.51 & 11.96 &  11.70 &0.298&0.60\\
12 26 37.32 & +22 34 53.4 & -12.0 &  -8.0 & 1.0 & 1.0 & 15.19 & 13.34 &  12.60 & 11.97 &  11.77 &0.288&0.60\\
12 33 30.31 & +28 12 55.9 & -12.0 & -14.0 & 2.0 & 11.0& 14.06 & 13.59 &  12.65 & 12.07 &  11.84 &0.279&0.60\\
12 16 29.21 & +23 32 32.9 & -12.0 &  -8.0 & 6.0 & 1.0 & 15.08 & 14.55 &  12.77 & 12.13 &  11.91 &0.270&0.60\\
12 30 46.17 & +23 45 49.0 & -10.0 &  -6.0 & 1.0 & 9.0 & 15.43 & 13.83 &  12.72 & 12.17 &  11.91 &0.269&0.60\\
12 19 37.99 & +26 34 44.7 &  -6.0 &  -8.0 & 4.0 & 2.0 & 16.46 & 14.03 &  12.78 & 12.24 &  11.92 &0.269&0.60\\
12 14 23.97 & +28 21 16.6 &  -4.0&   -8.0 & 2.0 & 1.0 & 15.52 &   0.0 &  12.83 & 12.14 &  11.92 &0.269&0.60\\
\hline
\end{tabular}
\end{center}
\end{table*}

\section{Results}

Our survey of the Coma Berenices open cluster has recovered 45 previously known members 
in total, 38 listed by Bounatiro and Arimoto (1992) and 7 unearthed by Odenkirchen et 
al (1998). Furthermore, it has identified 60 new candidate cluster members with magnitudes
down to K$_{\rm S}$=12. Beyond this magnitude, no statistically significant difference
between the cluster and the control samples was found, as the contamination by field stars 
is too great.  We believe that our survey is reasonably 
complete to this limit; we tried proper motion search radii of 7 and 5 mas yr$^{-1}$, but 
in both cases found increasing numbers of likely members were excluded. Expanding the search 
radius to greater than 10 mas yr$^{-1}$ unearthed no statistically significant increase in the
number of candidate members, as the candidates and control stars increased proportionally, leading to many candidates with extremely small probabilities of
being cluster members. Our survey is 
complete to 90\% at this radius, as explained earlier. As the stars get fainter however, the errors in their proper motion do increase -
which is to be expected. It is entirely possible that the completeness is less than 90\% for the faint stars.

We have estimated the masses of our candidates and the previously known cluster members 
using the evolutionary tracks of Girardi et al. (2002), for masses$\geq$ 1 $\Msun$, and the 
NEXTGEN models of Baraffe et al. (1998) for M$<$1 $\Msun$. 
The stars were binned according to K magnitude or J-K colour, for the vertical and diagonal portions of the main sequence. 
We then linearly interpolated between the model masses to estimate the masses of the cluster stars and our candidates.
These masses are shown in the final columns of Tables 1 and 2, and illustrated in Figure 5.   By 
multiplying the estimated masses of the stars by their membership probabilities (we assume
$P_{\rm membership}$=1.0 for previously known members) and 
summing, we determine a total cluster mass of $\sim$102$\Msun$. This in turn allows us 
to derive a tidal radius of 6.5 pc or 4.1$^{\circ}$ at 90 pc. Thus we find that within 
our adopted survey radius of 4$^{\circ}$ we should expect to find 99 per cent of the 
gravitationally bound proper motion selected cluster members. Indeed, increasing the search radius to 5$^{\circ}$,
led to a near equal increase in both candidate cluster members and stars in the control.

\begin{figure}
\scalebox{0.45}{{\includegraphics{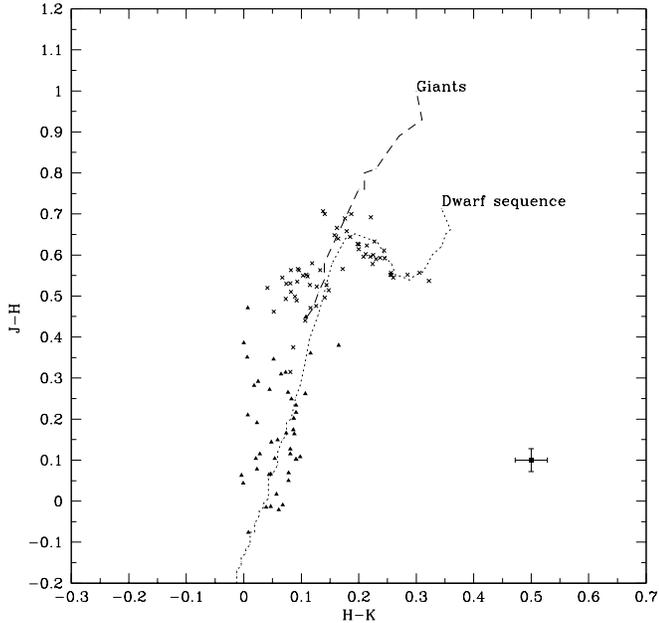}}}
\caption{A JHK colour-colour diagram for candidate (crosses) and previously known 
(solid triangles) cluster members. Empirical dwarf (dotted line) and giant (dashed line) 
sequences of Koornneef (1983), are overplotted. 
The diagram has been plotted using the 2MASS system. The colour transformations of Carpenter (2001) were used to 
convert the model colours.}
\end{figure}
\begin{figure}
\scalebox{0.45}{{\includegraphics{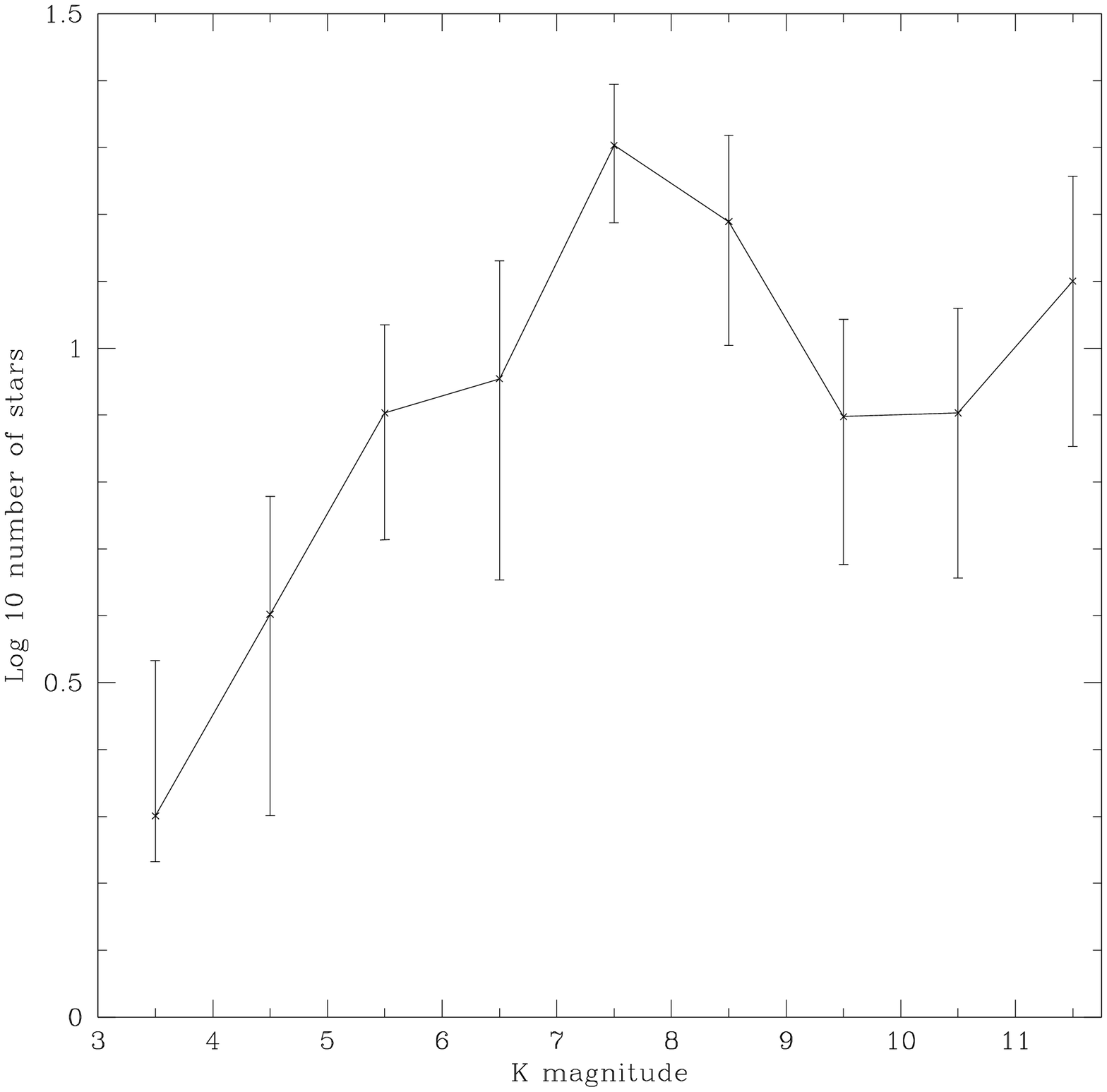}}}
\caption{Luminosity function for the cluster, taking into account probability of membership. The error bars indicate the Poisson error.}
\end{figure}
\begin{figure}
\scalebox{0.45}{{\includegraphics{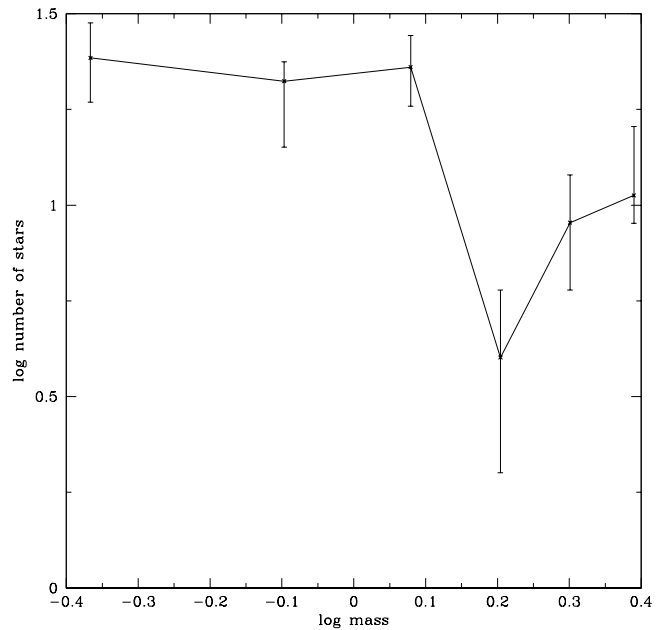}}}
\caption{Mass function for the cluster, taking into account probability of membership. The error bars indicate the Poisson error.}
\end{figure}

Figure 3 shows the J-H, H-K two colour diagram for all new candidate and previously 
known cluster members. 
The colours of our candidates show a clear turnover at H-K
$\approx$0.15, consistent with them being dwarfs and supporting our conclusion that 
these are low mass members of Melotte 111. Therefore, we have constructed a K$_{\rm S}$ luminosity 
function, and mass function for the cluster (Figures 4, 5).  The luminosty function 
 is seen to peak at K$_{\rm S}$ =7.5, M$_{K_{\rm S}}$=2.7 
and to gradually decline towards fainter magnitudes. Superimposed on 
this decline is the Wielen dip (Wielen 1971) at K$_{\rm S}$=10.5. 

The peaking of the luminosity function at this comparatively bright magnitude strongly 
suggests that a large proportion of the lower mass members, which are normally more 
numerous than the higher mass members (e.g. Hambly et al. 1991), have evaporated from 
the cluster. This should not come as a great surprise since at ages $\tau$$\simgreat$
200Myrs, dynamical evolution of a typical open cluster is expected to lead to the loss 
of a significant and increasing fraction of the lowest mass members (e.g. de la Fuente
 Marcos \& de la Fuente Marcos 2000). The mass function also supports this conclusion.
 Thus despite having found low mass stars in the
cluster our study does not contradict the conclusion of previous authors that the 
cluster is deficient in low mass stars. Odenkirchen et al. (1998) found that the 
luminosity function of extratidal stars associated with the cluster rises towards lower
masses. This can be considered to be indicative of the ongoing loss of low mass members.  
Ford et al (2001) however, determined that $\approx$ half of these stars did not belong to the 
cluster. A more in depth study of lower mass members would be required to determine a more 
accurate luminosity function at fainter magnitudes.

Our faintest candidate member has an estimated mass of 0.269 $\Msun$. There are some 
18 stars in the magnitude range K$_{\rm S}$=11-12, the lowest luminosity bin, thus it 
seems a distinct possibility that the Coma Berenices open cluster will have members 
right down to the limit of the hydrogen burning main sequence and possibly into the 
brown dwarf regime. This conclusion is also confimed by the rising mass function at lower masses.

\section{Conclusions}

We have performed a deep, wide area survey of the Coma Berenices open star cluster, using the USNO-B1.0 
and the 2MASS Point Source catalogues, to search for new candidate low mass members. This has led to the
identification of 60 objects with probabilities of cluster membership, $P_{\rm membership}$$\simgreat$0.6.
Our lowest mass new candidate member has M$\approx$0.269$\Msun$ in contrast to the previously known lowest
mass member, M$\approx$0.915$\Msun$. Thus we have extended considerably the cluster luminosity function 
towards the bottom of the main sequence. As reported by previous investigations of Melotte 111, the luminosity
function is observed to decline towards fainter magnitudes, indicating that the cluster has probably lost
and continues to loose its lowest mass members. This is not surprising for a  cluster whose age is 400-500Myrs.
Nevertheless, as the cluster luminosity function remains well above zero at K$_{\rm S}$=12, we believe the cluster 
probably has members down to the bottom of the hydrogen burning main sequence and possibly some brown dwarfs. Thus 
a deeper IZ survey of the cluster could prove a fruitful undertaking. 

\section{Acknowledgements}
This research has made use of the USNOFS Image and Catalogue Archive operated by the United States 
Naval Observatory, Flagstaff Station (http://www.nofs.navy.mil/data/fchpix/). This publication has 
also made use of data products from the Two Micron All Sky Survey, which is a joint project of the 
University of Massachusetts and the Infrared Processing and Analysis Center/California Institute 
of Technology, funded by the National Aeronautics and Space Administration and the National Science
Foundation. SLC and PDD acknowledge funding from PPARC.

\bsp

\label{lastpage}
{}


\begin{thebibliography}{}
\bibitem[\protect\citeauthoryear{Adams et al.}{2002}]{adams02} 
Adams J. ~D., Stauffer J.~R., Skrutskie, M.~F., Monet, D. ~ G., Portegies Zwart S. ~F., Janes K.~A.,
 Beichman C. ~A., 2002, AJ, 124, 1570

\bibitem[\protect\citeauthoryear{Argue \& Kenworthy}{1969}]{argue69} 
Argue A. ~N., Kenworthy C. ~K., 1969, MNRAS, 146, 479

\bibitem[\protect\citeauthoryear{Artyukhina}{1955}]{artyukhina55} 
Artyukhina, N. ~M., 1955, TrSht, 26, 3

\bibitem[\protect\citeauthoryear{Baraffe et al.}{1998}]{baraffe98} 
Baraffe I., Chabrier G., Allard F., Hauschildt P. ~H., 1998, A\&A, 337, 403

\bibitem[\protect\citeauthoryear{Bounatiro \& Arimoto}{1993}]{bounatiro93} 
Bounatiro L., Arimoto N., 1993, A\&A, 268, 829

\bibitem[\protect\citeauthoryear{Bouvier et al.}{1998}]{bouvier98} 
Bouvier J., Stauffer J. ~R., Martin E. ~L., Barrado y Navascues D., Wallace B., Bejar V. ~J. ~S., 1998, A\&A, 336, 490

\bibitem[\protect\citeauthoryear{Carpenter}{2001}]{carpenter01} 
Carpenter J. ~M., 2001, AJ, 121, 2851

\bibitem[\protect\citeauthoryear{Cayrel de Strobel}{1990}]{cayrel90} 
Cayrel de Strobel G., 1990, MnSAI, 61, 613

\bibitem[\protect\citeauthoryear{de la Fuente Marcos \& de la Fuente Marcos}{2000}]{delafuente00} 
de la Fuente Marcos R., de la Fuente Marcos C., 2000, Ap\&SS, 271, 127

\bibitem[\protect\citeauthoryear{Deluca \& Weis}{1981}]{deluca03} 
Deluca E. ~E., Weis E. ~W., 1981, PASP, 93, 32

\bibitem[\protect\citeauthoryear{Ford et al}{2001}]{ford01} 
Ford A., Jeffries R. ~D., James D. ~J., Barnes J. ~R., A\&A, 2001, 369, 871

\bibitem[\protect\citeauthoryear{Friel \& Boesgaard}{1992}]{friel92} 
Friel E. ~D., Boesgaard A. ~M., 1992, ApJ, 387, 107

\bibitem[\protect\citeauthoryear{Garcia-Lopez}{2000}]{garcialopez00} 
Garcia-Lopez R. ~J., Randich S., Zapatero Osorio M. ~R., Pallavicini R., 2000, A\&A, 363, 958

\bibitem[\protect\citeauthoryear{Girardi et al.}{2002}]{girardi02} 
Girardi L., Bertelli G., Bressan A., Chiosi C., Groenwegen M. ~A. ~T., Marigo P., Salasnich B., Weiss A., 2002, A\&A, 391, 195

\bibitem[\protect\citeauthoryear{Gizis et al.}{1999}]{gizis99} 
Gizis J. ~E., Reid I. ~N., Monet D. ~G., 1999, AJ, 118, 997

\bibitem[\protect\citeauthoryear{Hambly et al.}{1991}]{hambly91} 
Hambly N. ~C., Hawkins M. ~R. ~S., Jameson R. ~F., 1991, MNRAS, 253, 1

\bibitem[\protect\citeauthoryear{Jameson \& Skillen.}{1989}]{jameson89} 
Jameson R. ~F., Skillen I., 1989, MNRAS, 239, 247 

\bibitem[\protect\citeauthoryear{Koornneef}{1983}]{koorneef83} 
Koornneef J., 1983, A\&A, 128, 84

\bibitem[\protect\citeauthoryear{Lodieu et al.}{2005}]{lodieu05} 
Lodieu N., McCaughrean M. ~J., Barrado Y Navascus D., Bouvier J., Stauffer J. ~R., 2005, A\&A, 436, 853  

\bibitem[\protect\citeauthoryear{Madsen et al.}{2002}]{madsen02} 
Madsen S., Dravins D., Lindgren L., 2002, A\&A, 381, 446

\bibitem[\protect\citeauthoryear{Monet et al.}{2003}]{monet03} 
Monet D. ~G., et al., 2003, ApJ, 125, 984

\bibitem[\protect\citeauthoryear{Moraux et al.}{2001}]{moraux01} 
Moraux E., Bouvier J., Stauffer J. ~R., 2001, A\&A, 367, 211

\bibitem[\protect\citeauthoryear{Nicolet et al.}{1981}]{nicolet81} 
Nicolet B., 1981, A\&A, 104, 185

\bibitem[\protect\citeauthoryear{Odenkirchen et al.}{1998}]{odenkirchen98} 
Odenkrichen, M., Soubiran, C., Colin, J., 1998, New Astron., 3, 583

\bibitem[\protect\citeauthoryear{Oort}{1979}]{oort79} 
Oort J. ~H., 1979, A\&A, 78, 312

\bibitem[\protect\citeauthoryear{Perryman et al}{1998}]{perryman98} 
Perryman M. ~A. ~C., et al., 1998, A\&A, 331, 81

\bibitem[\protect\citeauthoryear{Randich}{1996}]{randich96} 
Randich S., Schmitt J. ~H. ~M. ~M., Prosser C., 1996, A\&A, 313, 815

\bibitem[\protect\citeauthoryear{Reid}{1992}]{reid92} 
Reid I. ~N., 1992, MNRAS, 257,257

\bibitem[\protect\citeauthoryear{Reid}{1993}]{reid93} 
Reid I. ~N., 1993, MNRAS, 265, 785

\bibitem[\protect\citeauthoryear{Skrutskie et al.}{1997}]{skrutskie97} 
Skrutskie M. ~F., et al ., 1997, in Garzon F. et al., eds, The Impact of Large Scale Near-IR Sky Surveys, Kluwer, Dordrecht, p.25 

\bibitem[\protect\citeauthoryear{Trumpler}{1938}]{trumpler38} 
Trumpler R. ~J., 1938, Lick Observatory Bull., 18, 167

\bibitem[\protect\citeauthoryear{Tsvetkov}{1989}]{tsvetkov89} 
Tsvetkov T. ~G., AP\&SS, 1989, 151, 47

\bibitem[\protect\citeauthoryear{van Leeuwen}{1980}]{vanleeuwen80} 
van Leeuwen F., 1980, in Star Clusters. Proc. Symp., Victoria, British Columbia. Canada, August 27-30, 1979. Reidel, Dordrecht, p. 157

\bibitem[\protect\citeauthoryear{van Leeuwen}{1999}]{vanleeuwen99} 
van Leeuwen F., 1999, A\&A, 341, L71

\bibitem[\protect\citeauthoryear{Wielen}{1971}]{wielen71} 
Wielen R.,  1971, A\&A, 13, 309

\bibitem[\protect\citeauthoryear{Zacharias et al.}{2004}]{zacharias04} 
Zacharias N., Urban S. ~E., Zacharias M. ~I., Wycoff G. ~L., Hall D. ~M., Monet D. ~G., Rafferty T. ~J., 2004, AJ, 127, 3043 


\end{thebibliography}
\end{document}